\documentclass[3p,times]{elsarticle}

\usepackage{ecrc}

\volume{00}

\firstpage{1}

\journalname{Procedia Computer Science}

\runauth{}

\jid{procs}

\jnltitlelogo{Procedia Computer Science}

\CopyrightLine{2011}{Published by Elsevier Ltd.}

\usepackage{amssymb}
\usepackage{amsmath}

\usepackage[figuresright]{rotating}

\begin{document}

\begin{frontmatter}



\dochead{}

\title{Improving Routing Efficiency through Intermediate Target Based Geographic Routing}


\author[label1]{Zongming Fei}
\address[label1]{{\bf Corresponding Author,} Department of Computer Science, University of Kentucky, Lexington,KY. Email: fei@netlab.uky.edu}
\author[label2]{Jianjun Yang}
 \address[label2]{Department of Computer Science and Information Systems, University of North Georgia, Oakwood, GA. Email: jianjun.yang@ung.edu}
\author[label3]{Hui Lu}
\address[label3]{School of Electronic and Information Engineering, Beihang University, \\
Beijing 100191, China, Email: luhui.buaa@gmail.com }


\address{}

\begin{abstract}
The greedy strategy of geographical routing may cause the local minimum problem
when there is a hole in the routing area. It depends on other strategies such
as perimeter routing to find a detour path, which can be long and result in
inefficiency of the routing protocol. In this paper, we propose a new approach
called Intermediate Target based Geographic Routing (ITGR) to solve the long
detour path problem. The basic idea is to use previous experience to determine
the destination areas that are shaded by the holes. The novelty of the approach
is that a single forwarding path can be used to determine a shaded area that
may cover many destination nodes. We design an efficient method for the source
to find out whether a destination node belongs to a shaded area. The source
then selects an intermediate node as the tentative target and greedily forwards
packets to it, which in turn forwards the packet to the final destination by
greedy routing. ITGR can combine multiple shaded areas to improve the
efficiency of representation and routing. We perform simulations and demonstrate
that ITGR significantly reduces the routing path length, compared with existing
geographic routing protocols.
\end{abstract}

\begin{keyword}
mobile ad hoc networks \sep greedy forwarding \sep location-based routing


\end{keyword}

\end{frontmatter}


\section{Introduction}
\label{intro}

In wireless networks, a node can communicate with a nearby neighbor node directly.
However, it is much more complicated when it needs to send messages to a destination
node farther away out of the range of its wireless signal. In this situation,
it relies on other nodes to relay its packets step by step until
they reach the destination.
Routing protocols~\cite{ex1,ex2,ex3,ex4,ex5,ex6} have been proposed to
find a routing path from a source node to a destination node.
They can be classified into proactive routing protocols and
on-demand routing protocols depending on when paths are determined.
Proactive protocols, such as DSDV~\cite{ex1}, TBRPF~\cite{ex2},
and OLSR~\cite{ex3}, exchange routing information periodically between hosts,
and constantly maintain a set of available
routes for all nodes in the network.
In contrast, on-demand (or reactive) routing protocols,
such as AODV~\cite{ex4}, DSR~\cite{ex5}, and TORA~\cite{ex6},
delay route discovery until a particular route is required, and propagate routing information only on demand.
There are also a few hybrid protocols, such as ZRP~\cite{ex7}, HARP~\cite{ex8}, and ZHLS~\cite{ex9},
which combine proactive and reactive routing strategies.
Most of these protocols involve broadcasting link state messages or
request messages in order to find a path. The flooding of information
can cause the scalability issue with these routing protocols.

Location information can be used to simplify the routing process in wireless networks.
Previous work has demonstrated that the location information can be obtained
either through GPS or by using virtual coordinates~\cite{ex10,ex11,ex12}.
Geographic routing exploits the location information and makes
the routing in ad hoc networks scalable.
The source node first acquires the
location of the destination node it wants to communicate with, then forwards the packet to one of its
neighbors that is closest to the destination.
This process is repeated until the packet reaches the destination.
A path is found via a series of independent local decisions rather than flooding.
Each node only maintains information about its neighbors.
However, geographic routing has to deal with the so-called local minimum phenomenon,
in which a packet may get stuck at a node that does not have a closer neighbor
to the destination, even though there is a path from the source to the destination
in the network.
This typically happens when there is a void area (or hole) that has no active nodes.
In wireless
ad hoc networks, the holes can be caused by various reasons~\cite{ex24}.
For instance, malicious nodes can jam
the communication to form jamming holes.
If the signal of nodes is not strong enough to cover everywhere in the network plane,
coverage holes may exist. Moreover, routing holes
can be formed either due to voids in node deployment or because of failure of nodes due to various reasons such as
malfunctioning, or battery depletion.

Many solutions have been proposed to deal with the local minimum problem.
Karp and Kung proposed the Greedy Perimeter Stateless
Routing (GPSR) protocol, which guarantees the delivery of the packet if a path exists~\cite{ex13}.
When a packet is stuck at a node, the protocol will route the packet around the faces of the graph
to get out of the local minimum. Several approaches were proposed
that are originated from the face routing. Although they can find the
available routing paths, they often cause the long detour paths. It is a hot topic
to avoid long detour path in the research community~\cite{ITGR,YangHDAR} and it has valued applications~\cite{vehicle1}.

To avoid such long detour paths, this paper proposes
a new approach called Intermediate Target Based Geographic Routing (ITGR).
The source
determines destination areas which are shaded by the holes based on
previous forwarding experience.
It also records one or more intermediate nodes called landmark nodes and
uses them as tentative targets. The routing path from the source node to the
next tentative target is greedy.
The routing paths from one tentative target to another and finally to the destination
are greedy as well. Hence the total routing path is constructed by a series of greedy routing paths.
The novelty of the approach is that
a single forwarding path can be used to determine an area that may cover many
destination nodes.
We design an efficient method for the source to find out whether a destination node belongs to
a shaded area.
Using
intermediate nodes as tentative targets and greedily forwarding packets to them can
avoid the original long detour paths.
To further improve the efficiency of representation and routing, we
design the mechanism for ITGR to combine multiple shaded areas.
Simulations show that ITGR reduces routing path length by 17\%  and
the number of forwarding hops by 15\%, compared with GPSR.

The rest of the paper is organized as follows.
Section~\ref{related} discusses related work on geographical routing and how the
local minimum problem is dealt with.
Section~\ref{shadedarea} proposes a novel method for detecting shaded areas and
presents a new Intermediate Target based Geographic Routing protocol.
It also presents the method for combining multiple cache entries to
save the state information and reduce the search time.
Section~\ref{evaluation}
evaluates the proposed schemes by simulations and describes performance results.
Section~\ref{conclusion} concludes the paper.

\section{Related Work}
\label{related}

Many geographic routing protocols have been developed for ad hoc networks.
In early protocols,
each intermediate node in the network forwards packets to its neighbor closest to
the destination, till the destination is
reached.
Packets are simply dropped when greedy
forwarding causes them to end up at a local minimum node.

To solve the local minimum problem, geometric face routing algorithm
(called Compass routing)~\cite{ex16} was proposed that guarantees packet delivery
in most (but not all) networks.
Several practical algorithms, which are variations of face routing,  have since been developed.
By combining greedy and face routing, Karp and Kung proposed the Greedy Perimeter Stateless Routing
(GPSR) algorithm~\cite{ex13}.
It consists of the greedy forwarding mode and the perimeter forwarding mode,
which is applied in the regions where the greedy forwarding does not work.
An enhanced algorithm, called Adaptive Face
Routing (AFR), uses an ellipse to restrict the search area
during routing so that in the worst case,
the total routing cost is no worse than a constant factor of the cost for the optimal route~\cite{ex17}.
The latest addition to the face routing related family is GPVFR,
which improves routing efficiency by exploiting local
face information~\cite{ex18}.

To support geometric routing better in large wireless networks,
several schemes were proposed to maintain geographic information on planar faces~\cite{ex19}.
Gabriel Graph~\cite{gg} and Relative Neighborhood Graph~\cite{rng} are earlier sparse planar graphs
constructed by planarization algorithms,
with the assumption that the original graph is a unit-disk graph (UNG)~\cite{udg}.
Dense planar graphs are constructed from UNGs based on Delaunay triangulation~\cite{PLDel}.
The Cross-Link Detection Protocol (CLDP)~\cite{ex19} produces a subgraph on which face-routing-based algorithms
are guaranteed to work correctly without making a unit-disk graph assumption.
The key insight is that starting from a connected graph, nodes can
independently probe each of their links using a right-hand rule to determine
if the link crosses some other link in the network.
%

More recently, an idea based on the method of figuring out the void areas
in advance was explored.
A node keeps the coordinates of key nodes as well as the locations of its neighbors.
The forwarding nodes will use the information to avoid approaching the
holes~\cite{ex20,ex21,ex22}.
Also related is GLR,
a geographic routing scheme for large wireless ad hoc networks~\cite{ex26}.
In the algorithm, once a source node sends packets to a destination
node and meets a hole, the source node saves the location of the landmark node
to its local cache. If any packet is to be forwarded to the {\em same} destination,
the source node will forward
the packet through the landmark.
So each entry in the cache can only be used for a single destination node.
In contrast, our approach learns from previous experience and generalizes it to cover an area
of destination nodes. The number of nodes that can benefit from one cache entry can be
orders of magnitude larger.
%
Yet we design a simple way to represent the area and
an efficient algorithm to decide whether a destination node is in the area.


\section{Intermediate Target Based Routing}
\label{shadedarea}

\subsection{The Basic Idea}

\begin{figure}[hbt]
\centering
\includegraphics[width=8.0cm]{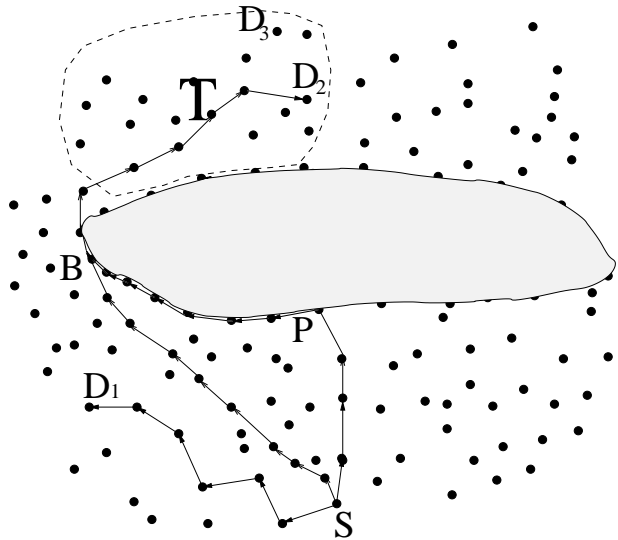}
\caption{Greedy path and detour path}
\label{fig1}
\end{figure}

We use a simple example to illustrate the basic idea of our approach.
We assume that all nodes are static and distributed in a two dimensional space.
%
As shown in Fig.~\ref{fig1}, we assume that
$S$ is the source node and
$D_1$, $D_2$ and $D_3$ are three different destination nodes.
When $S$ wants to send packets to $D_1$, it can find an efficient path by greedy forwarding.

However,
when $S$ wants to send a packet to $D_2$, it uses the greedy forwarding and the packet will reach node $P$.
Because of the existence of the void area, $P$ is closer to $D_2$ than all of the $P$'s neighbors.
So $P$ cannot reach $D_2$ by greedy forwarding and is called a {\em local minimum node}.
Fortunately, we have various routing algorithms~\cite{ex13} to let
$P$ change from the greedy mode to the perimeter routing mode.
The packet will be forwarded along a
detour path until it arrives at node $B$,
where the forwarding mode is changed from the perimeter routing mode
to greedy forwarding.
Node $B$ is called a {\em landmark node}.
After node $B$, the packet can be forwarded to destination $D_2$ by greedy forwarding.
Because $D_2$ is shaded by the hole, the original simple greedy forwarding
has to take a detour. This detour path can be long.

To deal with routing inefficiency caused by the detour,
we can let either destination node $D_2$ or landmark node $B$ inform source $S$
that such a detour occurred.
After receiving the message, $S$ keeps a record that associates $D_2$ with $B$,
meaning that if
the destination is $D_2$, forward through intermediate node $B$.
After that, if $S$ later needs to send packets to $D_2$, it
can send them to $B$ first (using $B$ as an intermediate target) by greedy forwarding.
The path will be from $S$ to $B$ and then to $D_2$, instead of from $S$ to $P$, to $B$, and
then to $D_2$. This new path can be much shorter and may be the best path to get to $D_2$ from $S$.
The significance of the technique depends on how likely $S$ needs to send packets to $D_2$ again.

Now consider that $S$ needs to send a packet to $D_3$. Most likely, it will be forwarded to $P$
by greedy forwarding,
then go through a detour using perimeter routing to $B$, and finally reach $D_3$.
The question we are interested in is whether the detour information about $D_2$ can be used to guide
the forwarding by $S$ for packets to $D_3$. In another word, can we generalize the strategy of
using the intermediate node $B$ for packet forwarding from the single
destination node $D_2$ to multiple nodes?

The basic idea of this paper is to find a shaded area $T$ such that for any destination node $D \in T$,
source node $S$ can benefit from using $B$ as an intermediate target. Packets will be forwarded from $S$
to $B$ using greedy forwarding and then $B$ will relay the packets to the final destination
using greedy forwarding. The challenge is to find a simple representation of
shaded area $T$ and an efficient algorithm to determine whether a target node $D$ is
in the shaded area.



\subsection{Shaded Area}


The shaded area for source node $S$ can be determined by the
locations of
the local minimum node $P$, the landmark node $B$ and the source node $S$.
This is a learning process for $S$  when it finds out that its packets are
sent over a detour path.
When a packet arrives at a node in perimeter mode, this node will determine whether it is a landmark node
by checking whether it should change the forwarding mode to Greedy. If it is a landmark node, it will inform
the source node of its own location ($B$) and the location of the local minimum node $P$ (recorded in
the packet).

\begin{figure}[!htp]
\begin{center}
\includegraphics[width=8.0cm]{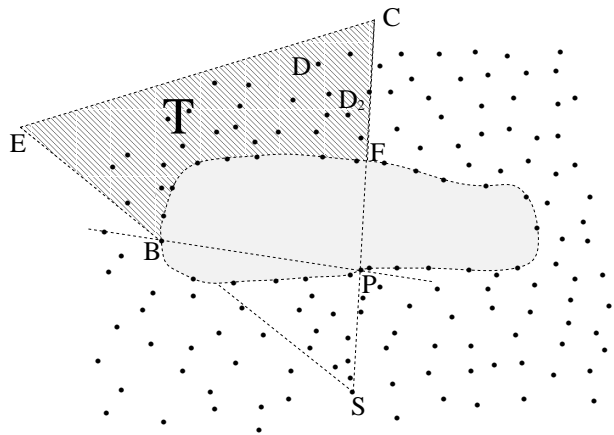}
\end{center}
\caption{The shaded area}
\label{fig2}
\end{figure}

When $S$ learns the locations of $B$ and $P$, we can define a shaded area as shown in Fig.~\ref{fig2}.
%
%
%
We connect $S$ with $P$ using a straight line and extend it to intersect with the hole at another point $F$.
Ray $SF$ further extends to some point $C$.
We connect $S$ with $B$ using a straight line and extend it some point $E$.
Then the area semi-enclosed by $EB$, the perimeter from $B$ to $F$ and $FC$ is the
shaded destination area $T$.
Hence, if $S$ needs to send packets to any destination node $D$ in $T$,
the destination is hidden behind the hole.
To avoid a detour path, $S$ sends the packets to $B$ first, and $B$ will then relay them to $D$.
Both paths can be greedy paths.
We observe that for some destination node $D' \in T$, the greedy forwarding from $S$ to $D'$ may
be stuck at a different local minimum node (other than $P$). However, forwarding to $B$ first
can still benefit by having a shorter path than going through the local minimum node.

Given a destination node $D$, we need to determine whether it is in the destination area $T$.
As shown in Fig.~\ref{fig2}, the area is enclosed by rays and partial edges of the hole polygon.
To simplify the calculation, our first step is to extend the shaded area to include
the area enclosed by line $BP$, line $PF$ and arc $BF$,
since it is in the void area and has no active nodes.
The new destination area becomes the area semi-enclosed by $EBPC$.

After this extension, the determination of a destination node $D$ in shaded area becomes simple.
If a destination node $D$ satisfies the following conditions, it must be located in the shaded area. \\

\text{} 1) $D$ and $P$ are located on the same side of line $SB$;\\
\text{~~~~~~} 2) $D$ and $B$ are located on the same side of line $SP$; and \\
\text{~~~~~~} 3) $D$ and $S$ are located on the opposite sides of line $BP$.

\vspace{0.1in}

Suppose that the coordinates of nodes $S$, $B$ and $P$ are $S(x_s,y_s)$, $B(x_b,y_b)$ and $P(x_p,y_p)$,
respectively.
Line $SB$ can be described by the following equation.
\begin{displaymath}
\frac{y-y_s}{y_b-y_s}=\frac{x-x_s}{x_b-x_s}.
\end{displaymath}

It can be written as
\begin{equation}
(y_b-y_s)x-(x_b-x_s)y+(x_by_s-x_sy_b)=0.
\label{eq1}
\end{equation}

Let $f_1(x,y) = (y_b-y_s)x-(x_b-x_s)y+(x_by_s-x_sy_b)$.
Suppose D's coordinates are $D$$(x_d,y_d)$.
$D$ and $P$ are located on the same side of line $SB$ if and only if
$f_1(x_d,y_d) * f_1(x_p, y_p) > 0$. To include
the case of $D$ being on line $SB$, we can use
\begin{equation}
f_1(x_d,y_d) * f_1(x_p, y_p) \ge 0.
\label{eq2}
\end{equation}

Similarly, we can find the equation for line SP as
\begin{equation}
f_2(x,y) = (y_p-y_s)x-(x_p-x_s)y+(x_b y_s - x_s y_p) = 0,
\label{eq3}
\end{equation}

and the equation for line $BP$ as
\begin{equation}
f_3(x,y) = (y_b-y_p)x-(x_b-x_p)y+(x_b y_p-x_p y_b) = 0.
\label{eq4}
\end{equation}

Nodes $D$ and $B$ are located on the same side of line $SP$
if
\begin{equation}
f_2(x_d,y_d) * f_2(x_b, y_b) \ge 0.
\label{eq5}
\end{equation}

Nodes $D$ and $S$ are located on the opposite sides of line $SP$
if
\begin{equation}
f_3(x_d,y_d) * f_3(x_s, y_s) \le 0.
\label{eq6}
\end{equation}

If all three conditions~(\ref{eq2}), (\ref{eq5}) and (\ref{eq6}) are met, node $D$ is in the shaded area.

\subsection{ITGR Routing Scheme}

In ITGR routing,
besides the source address $S$ and the destination address $D$,
a packet may contain
a list of intermediate targets $<I_1, I_2, \cdots, I_k>$,
which will be called {\em ITGR list} for the rest of the paper.
We define {\em the target ${\cal T}$ of a packet} as either the first element on the
ITGR list if the list exists, or the destination address $D$ if the list
does not exist.
Similar to other geographic routing schemes, a packet
forwarded in ITGR routing can be either in
Greedy mode or perimeter mode.
Theoretically it can
use any perimeter routing algorithm. However, for simplicity of
presentation, we assume that GPSR is used. Therefore, perimeter mode will also be
called {\em GPSR mode}.
As stated in GPSR routing, packets in GPSR mode will contain the location
of the local minimum node $P$, at which forwarding is changed
from Greedy mode to GPSR mode.

In ITGR routing, nodes have a local cache with
entries representing {\em shaded areas}.
Each shaded area is in the form of
$<P_i, B_i>$, where $P_i$ is
the location of the local minimum node and $B_i$
is the location of the landmark node.

\begin{figure}[hbt]
\baselineskip=10.2pt
{\small
  \noindent\underline{\bf ITGR\_send()} \\
  \text{~~~~}\textbf{if} the packet contains the ITGR list \\
  \text{~~~~~~~~}${\cal T}$ = first element of the ITGR list; \\
  \text{~~~~}\textbf{else} ${\cal T} = D$; \\
  \text{~~~~}Search local cache; \\
  \text{~~~~}\textbf{if} ${\cal T}$ is in a shaded area \\
  \text{~~~~~~~~}Extract the list of landmark nodes $<B_k, \cdots,B_1>$; \\
  \text{~~~~~~~~}\textbf{if} ITGR list exists \\
  \text{~~~~~~~~~~~~}Prepend $<B_k, \cdots,B_1>$ to the list; \\
  \text{~~~~~~~~}\textbf{else} Create ITGR list $<B_k, \cdots,B_1>$;\\
  \text{~~~~~~~~}${\cal T} = B_k$; \\
  \text{~~~~}\textbf{if} there is a neighbor closer to ${\cal T}$; \\
  \text{~~~~~~~~}Greedy mode forwarding to the neighbor closest to ${\cal T}$; \\
  \text{~~~~}\textbf{else}\\
  \text{~~~~~~~~}Record the current node as the local minimum node in the packet; \\
  \text{~~~~~~~~}GPSR mode forwarding; \\
}

\protect\caption{ITGR Sending Algorithm}
\protect\label{routingfig}
\end{figure}

When source $S$ needs to send a packet to destination $D$, it
calls function $ITGR\_send()$.
As described in Fig.~\ref{routingfig},
$ITGR\_send()$ first gets the target ${\cal T}$ of the packet.
It
searches its local cache to see
whether target ${\cal T}$ is in any of the shaded areas.
If yes, it extracts the landmark node ($B_1$).
Use this landmark node $B_1$ as the destination and search whether it is
in any shaded area. If it is, we get landmark node $B_2$.
This process will continue until we have a landmark node $B_k$
not in any shaded area.
Assume the list of landmark nodes we get is $B_1, B_2, \cdots, B_k$.
If the packet does not contain an ITGR list, it creates one
with elements
$B_k, B_{k-1}, \cdots, B_1$.
If the packet has an ITGR list, these elements are added in
the front.
We expect that in most cases, this list contains only one element $B_1$.
After that, we need to set ${\cal T}$ to the value of the first element
of the ITGR list.

As a last step, it forwards the packet to the neighbor
that is closest to ${\cal T}$.
If no neighbor is closer to ${\cal T}$ than the current node,
it changes the packet to GPSR mode
and follow the GPSR rules for forwarding (including putting
the address of the current node as the local minimum node in the packet.)
Note that $ITGR\_send()$ is not only used by the original source node,
but will be used by other intermediate nodes along the path.
In that case, it is called by the $ITGR\_process()$ function described in Fig.~\ref{receivingfig}.
The change from Greedy to GPSR mode is more likely to happen
at those intermediate nodes than the original source node.


After a node receives a packet from a neighbor, it will process the packet.
The node has to deal with several cases. It can be the final destination
node, the intermediate target node, the local minimum node,
the landmark node, or other forwarding nodes in the path.
In most cases, the node will call $ITGR\_send()$ to forward the packet
to the next hop.

\begin{figure}[hbt]
\baselineskip=10.2pt

{\small
  \noindent\underline{\bf ITGR\_process()} \\
  \text{~~~~}\textbf{if} its address is equal to destination $D$ \\
  \text{~~~~~~~~}Forwarding is finished and exit;\\
  \text{~~~~}\textbf{if} ITGR list exists and its address is equal \\
  \text{~~~~~~~~~~~~}to the first element of ITGR list \\
  \text{~~~~~~~~}Remove its address from the list; \\
  \text{~~~~~~~~}Call ITGR\_send() to send the packet to next hop; \\
  \text{~~~~}\textbf{elseif} the packet is in Greedy mode forwarding \\
  \text{~~~~~~~~}Call ITGR\_send() to send the packet to next hop; \\
  \text{~~~~}\textbf{elseif} the packet is in GPSR mode forwarding \\
  \text{~~~~~~~~}Set the value of ${\cal T}$ as the target of the packet;\\
  \text{~~~~~~~~}\textbf{if} the current node has a neighbor closer to ${\cal T}$\\
  \text{~~~~~~~~~~~~}Send a $landmark\_exist\_msg$ to source $S$ with \\
  \text{~~~~~~~~~~~~~~~~}local minimum node $P$ and its own address \\
  \text{~~~~~~~~~~~~~~~~}as the landmark node; \\
  \text{~~~~~~~~~~~~}Change to Greedy mode forwarding and call ITGR\_send(); \\
  \text{~~~~~~~~}\textbf{else} Continue GPSR forwarding; \\
}

\protect\caption{ITGR Processing Algorithm}
\protect\label{receivingfig}
\end{figure}

Fig.~\ref{receivingfig} describes processing algorithm $ITGR\_process()$
that a node will run after receiving a packet.
It first checks whether its address is equal to
destination $D$. If it is, the forwarding process is finished.
Otherwise, it checks whether there is an ITGR list and whether
its address is equal to the first element on the list.
If that is the case, it is the intermediate target.
Thus it removes itself from the list
and then calls $ITGR\_send()$ to send the packet to the next hop.

Next, depending on the forwarding mode of the packet, $ITGR\_process()$
processes the packet differently.
If the packet is in Greedy mode, the algorithm calls $ITGR\_send()$
to forward the packet to the next hop.
If the packet is in GPSR mode, the algorithm will do GPSR processing.
Specifically, if the condition of changing to Greedy mode is satisfied according to
GPSR routing,~\footnote{Such condition can be that the the forwarding
node finds out that one of its neighbors is closer to D than itself.}
it will change the forwarding mode to Greedy.
In addition to forwarding the packet by calling $ITGR\_send()$, it sends a
$landmark\_exist\_msg$ to source $S$ with the locations of the local minimum
node $P$ and its own (as the landmark).
Otherwise, it continues GPSR forwarding.

%
%
%
%
%
%

When the source receives
$landmark\_exist\_msg$,
it will put the local minimum node $P$ and the landmark node $B$ as an entry in its local cache.

\subsection{Combining Entries about Shaded Areas}
\label{merge}

If node $S$ sends many packets to different destinations, several detour paths will be generated by the GPSR routing strategy.
In this way, multiple entries with the format $<$ $LocalMinimum$, $Landmark$ $>$ might be
generated and saved in the cache of node $S$.
Among these entries, some shaded areas may overlap with each other. They
can have the same or different landmark nodes.
Though these cache entries can be used in their original form,
however, merging them can save space and facilitate efficient entry lookup.
In this section, we investigate how multiple entries in the cache can be combined.

Once node $S$ receives a $landmark\_exist\_msg$
$<P, B>$ from a landmark node, instead of inserting the new entry into the cache
directly, it first
looks up the entries in its local cache and possibly combines the new entry with an existing entry.
There are two situations $S$ needs to handle. One is that $S$ finds an existing entry in its cache with the same landmark $B$.
The other is that
$S$ finds an existing entry in its cache whose landmark is not $B$,
but the corresponding shaded area overlaps with the shaded area of $<P, B>$.

In the first situation, suppose that $S$ finds an entry $<$ $P^{'}$, $B$ $>$ existing in its cache. $S$ then updates
its entries as follows.~\footnote{Note that $<$ $P$, $B$ $>$ also represents the area determined by
the entry $<$ $P$, $B$ $>$.}\\*

\begin{figure}[!htp]
\begin{center}
\includegraphics[width=8.0cm]{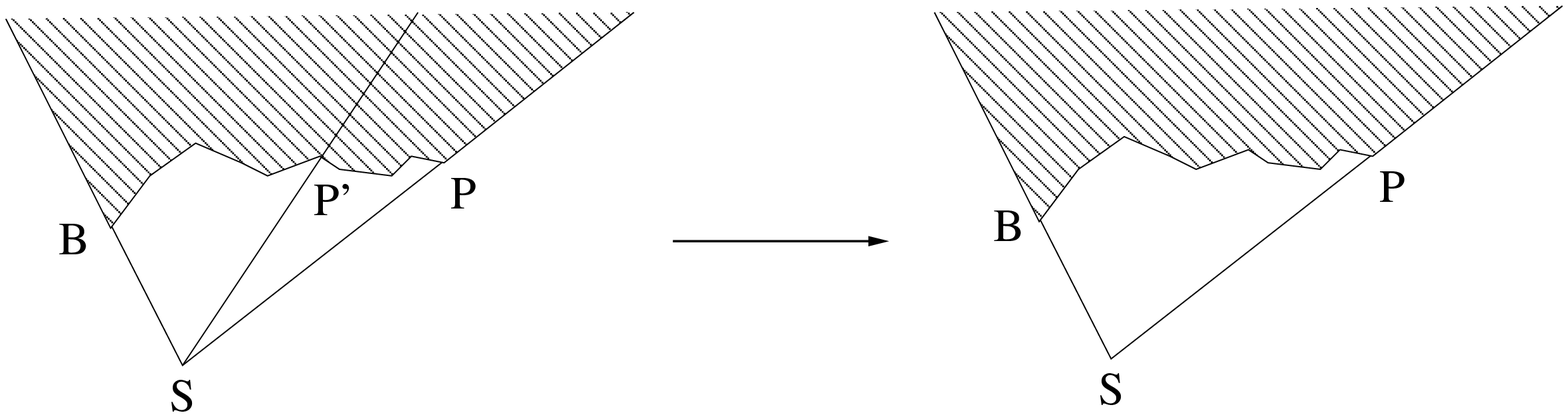}
\end{center}
\caption{Entry update: Case 1.}
\label{fig3_6}
\end{figure}

Case 1: $<$ $P^{'}$, $B$ $>$ $\subset$ $<$ $P$, $B$ $>$. This is the case in which $B$ and $P$ are on the opposite sides of $SP^{'}$
(Fig.~\ref{fig3_6}). This scenario can be determined by the coordinates of these points as follows. Suppose
the coordinates of points $S$, $B$, $P$ and $P^{'}$  are $S(x_s,y_s)$, $B(x_b,y_b)$,
$P(x_p,y_p)$ and $P^{'}(x_{p^{'}},y_{p^{'}})$, respectively. Then the equation of line $SP^{'}$  is:
\[(y_{p^{'}}-y_s)x-(x_{p^{'}}-x_s)y+(x_{p^{'}}y_s-x_sy_{p^{'}})=0. \]

Let $g_1(x,y) = (y_{p^{'}}-y_s)x-(x_{p^{'}}-x_s)y+(x_{p^{'}}y_s-x_sy_{p^{'}})$.
Nodes $B$ and $P$ are located on the opposite sides of line $SP^{'}$
if
\begin{equation}
g_1(x_b,y_b) * g_1(x_p, y_p) \le 0.
\label{eqm}
\end{equation}
$S$ updates the entries by removing $<$ $P^{'}$, $B$ $>$ and inserting $<$ $P$, $B$ $>$. \\*

\begin{figure}[!htp]
\begin{center}
\includegraphics[width=8.0cm]{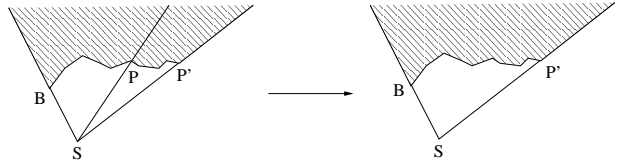}
\end{center}
\caption{Entry update: Case 2.}
\label{fig3_7}
\end{figure}

Case2: $<$ $P$, $B$ $>$  $\subset$  $<$ $P^{'}$, $B$ $>$. This is the case in which $B$ and $P^{'}$ are on the opposite sides of $SP$.
(Fig.~\ref{fig3_7}). We can also use coordinates of the points and the equation of line $SP$ to determine their relative locations.
Because the existing entry $<$ $P^{'}$, $B$ $>$
covers the new entry $<$ $P$, $B$ $>$, $S$ simply discards $<$ $P$, $B$ $>$ . \\*

The second situation is that  $S$ finds a new entry
$<$ $P'$, $B'$ $>$ related with $<P, B>$,
but they have two different landmarks $B$ and $B'$.
We discuss different scenarios in which their corresponding shaded areas overlap with each other.
Otherwise, $S$ can simply insert the new entry.

\begin{figure}[!htp]
\begin{center}
\includegraphics[width=3.5cm]{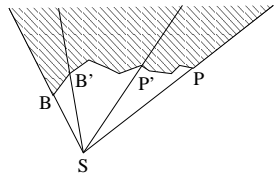}
\end{center}
\caption{Combining entries: Case 1.}
\label{fig3_11}
\end{figure}

Case 1: $<$ $P^{'}$, $B^{'}$ $>$ $\subset$ $<$ $P$, $B$ $>$.
This is the case in which $B$ and $P$ are on the opposite sides of $SB^{'}$,
and $B$ and $P$ are on the opposite sides of $SP^{'}$ (Fig.~\ref{fig3_11}).
The update is that
$S$  removes $<$ $P^{'}$, $B^{'}$ $>$ and then
   inserts $<$ $P$, $B$ $>$.
$S$ does this update because $<$ $P$, $B$ $>$ fully covers $<$ $P^{'}$, $B^{'}$ $>$.

\begin{figure}[!htp]
\begin{center}
\includegraphics[width=3.5cm]{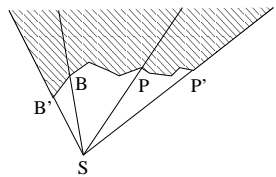}
\end{center}
\caption{Combining entries: Case 2.}
\label{fig3_12}
\end{figure}

Case 2: $<$ $P$, $B$ $>$ $\subset$ $<$ $P^{'}$, $B^{'}$ $>$.
This is the case in which $B$ and $P$ are on the same side of $SB^{'}$,
and also on the same side of $SP^{'}$ (Fig.~\ref{fig3_12}).
In this scenario, $S$ discards $<$ $P$, $B$ $>$ because the area determined by the new entry $<$ $P$, $B$ $>$
is covered by the existing entry $<$ $P^{'}$, $B^{'}$ $>$.

\begin{figure}[!htp]
\begin{center}
\includegraphics[width=3.5cm]{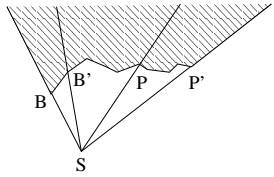}
\end{center}
\caption{Combining entries: Case 3.}
\label{fig3_13}
\end{figure}

Case 3: $<$ $P$, $B$ $>$ and $<$ $P^{'}$, $B^{'}$ $>$ are overlapped as follows. $B$ and $P$ are on the opposite sides of $SB^{'}$,
and $B$ and $P$ are on the same side of $SP^{'}$ (Fig.~\ref{fig3_13}).
The update is that
$S$ keeps the entry $<$ $P^{'}$, $B^{'}$ $>$ and inserts a new entry $<$ $B^{'}$, $B$ $>$. $S$ does this because
the new entry $<$ $P$, $B$ $>$ can be considered as two area $BSB^{'}$ and $B^{'}SP$. $B^{'}SP$ is included in $B^{'}SP^{'}$,
so only $BSB^{'}$ is inserted.

\begin{figure}[!htp]
\begin{center}
\includegraphics[width=3.5cm]{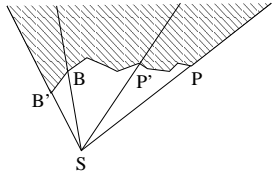}
\end{center}
\caption{Combining entries: Case 4.}
\label{fig3_14}
\end{figure}

Case 4: $<$ $P$, $B$ $>$ and $<$ $P^{'}$, $B^{'}$ $>$ are overlapped as follows. $B$ and $P$ are on the
same side of $SB'$,
and $B$ and $P$ are on the opposite sides of $SP'$ (Fig.~\ref{fig3_14}).
The update is that
$S$ removes the entry $<$ $P^{'}$, $B^{'}$ $>$ and then inserts two new entries $<$ $B$, $B^{'}$ $>$ and $<$ $P$, $B$ $>$.

\section{Performance Evaluation}
\label{evaluation}

We use the easim3D wireless network simulator~\cite{ex27} to evaluate the performance of the proposed mechanism.
We use a noiseless immobile radio network environment with an area of 400mX400m.
Nodes distributed in the area have a transmission radius of 40 meters.

We implemented both the GPSR routing protocol and our ITGR routing protocol using this simulation model.
Two metrics, the length of routing path and the number of hops, are used.
The number of nodes (density) varies from 50 to 300
with an increment of 50. For each case,
10 connected networks are generated with void areas set
inside the network.

\begin{figure}[hb]
\begin{center}
\includegraphics[width=9.0cm]{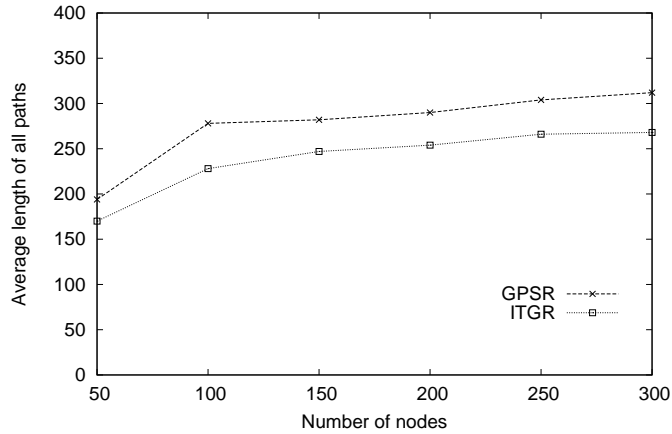}
\end{center}
\caption{The average path length.}
\label{fig15}
\end{figure}

\begin{figure}
\begin{center}
\includegraphics[width=9.0cm]{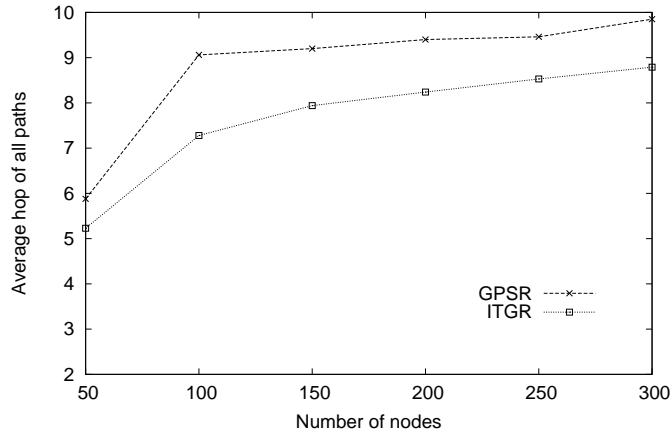}
\end{center}
\caption{The average number of hops.}
\label{fig16}
\end{figure}

Fig.~\ref{fig15} shows the average path length when the number of nodes changes from 50 to 300.
The average path length in ITGR is 17.52\% shorter than that of GPSR when there are 50 nodes
in the network.
When the density of networks increases, the ITGR performs a little bit better.
Fig.~\ref{fig16} shows the average number of hops with the number of nodes changing
from 50 to 300.
Similarly, the average number of hops in ITGR is 14.97\% less than that of GPSR
in the 50 node case.
In both Fig.~\ref{fig15} and Fig.~\ref{fig16}, the path length and the hop count
with 50 nodes (both GPSR and ITGR) are much smaller than other cases.
This is because on the network plane, to guarantee the network's connectivity, 50 nodes have to be
distributed in a relatively smaller area. This results in the shorter length and smaller number
of hops.

To further illustrate the effect of ITGR on path length and hop count, we divide the tested paths into two types. For a
routing path in ITGR routing, if no node in this path uses ITGR list for routing,
we call this path a type 1 path. Otherwise the path is a type 2 path.
We collect the data for the paths when GPSR routing is used.



\begin{table}[htp]
\caption{The average  percentage of type 2 paths over all paths}
\centering          
\begin{tabular}{|c| c| c| c| c| c| c|}    
\hline                        
The number of nodes & 50 & 100 & 150 & 200 & 250 & 300 \\ [0.5ex]  
\hline                     
Percentage &23.2&21.1&18.7&17.6&16.4&16.2 \\      
[1ex]        
\hline          
\end{tabular}
\label{table2}    
\end{table}

The percentage of type 2
path over all paths  is shown in Table~\ref{table2}.
It ranges from 23.2\% for 50 node networks and 16.2\% for 300 node networks.
The larger the number of nodes in the network, the smaller the percentage.
This can be explained as follows.
In the simulations, the nodes are distributed in a plane with a fixed size.
The size of holes in sparse networks is larger than
that in dense networks.
Therefore, more paths are affected by void areas when the number of nodes is small.

\begin{figure}[!htp]
\begin{center}
\includegraphics[width=9.0cm]{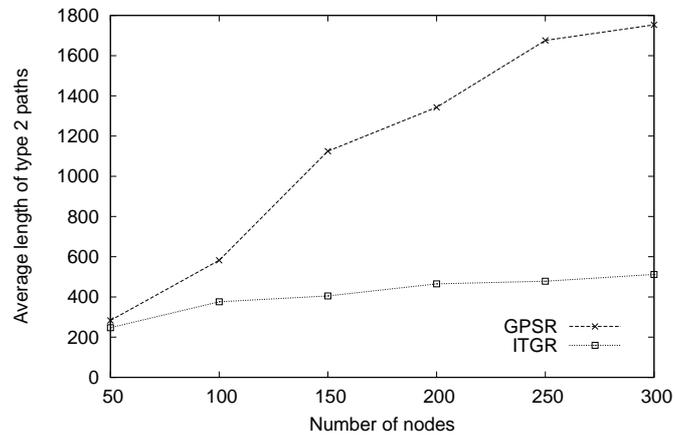}
\end{center}
\caption{The average length of type 2 paths.}
\label{fig19}
\end{figure}

\begin{figure}[htp]
\begin{center}
\includegraphics[width=9.0cm]{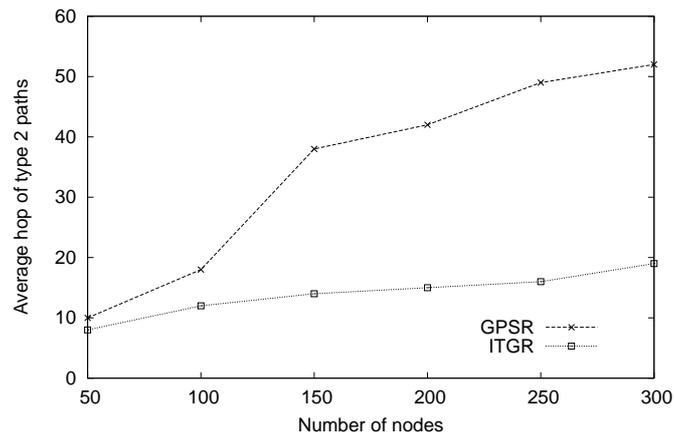}
\end{center}
\caption{The average number of hops of type 2 paths.}
\label{fig20}
\end{figure}

Fig.~\ref{fig19} and  Fig.~\ref{fig20} compare the performance of type 2 paths only.
Compared with GPSR,
ITGR has much shorter paths and fewer hops.
The gap between  ITGR and GPSR increases when the number of nodes in networks increases.
This is because when the number of nodes is larger,
detour paths generated by GPSR are longer.
For type 2 paths, the average length of ITGR is only 29.5\%
that of GPSR and the number of hops is only 27.3\% for 300 node networks.
From these two figures, we can see that ITGR shortens the long paths significantly.

\begin{figure}[!htp]
\begin{center}
\includegraphics[width=9.0cm]{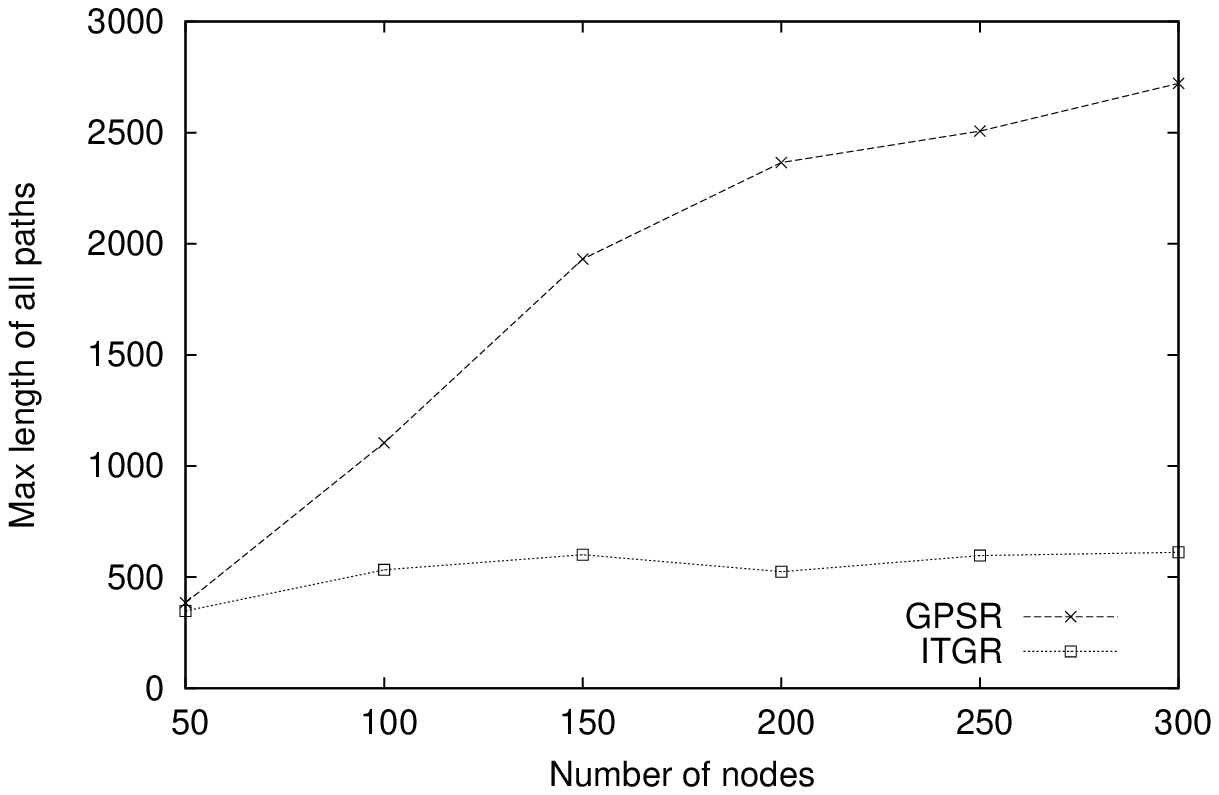}
\end{center}
\caption{The length of longest paths}
\label{fig21}
\end{figure}

\begin{figure}[!htp]
\begin{center}
\includegraphics[width=9.0cm]{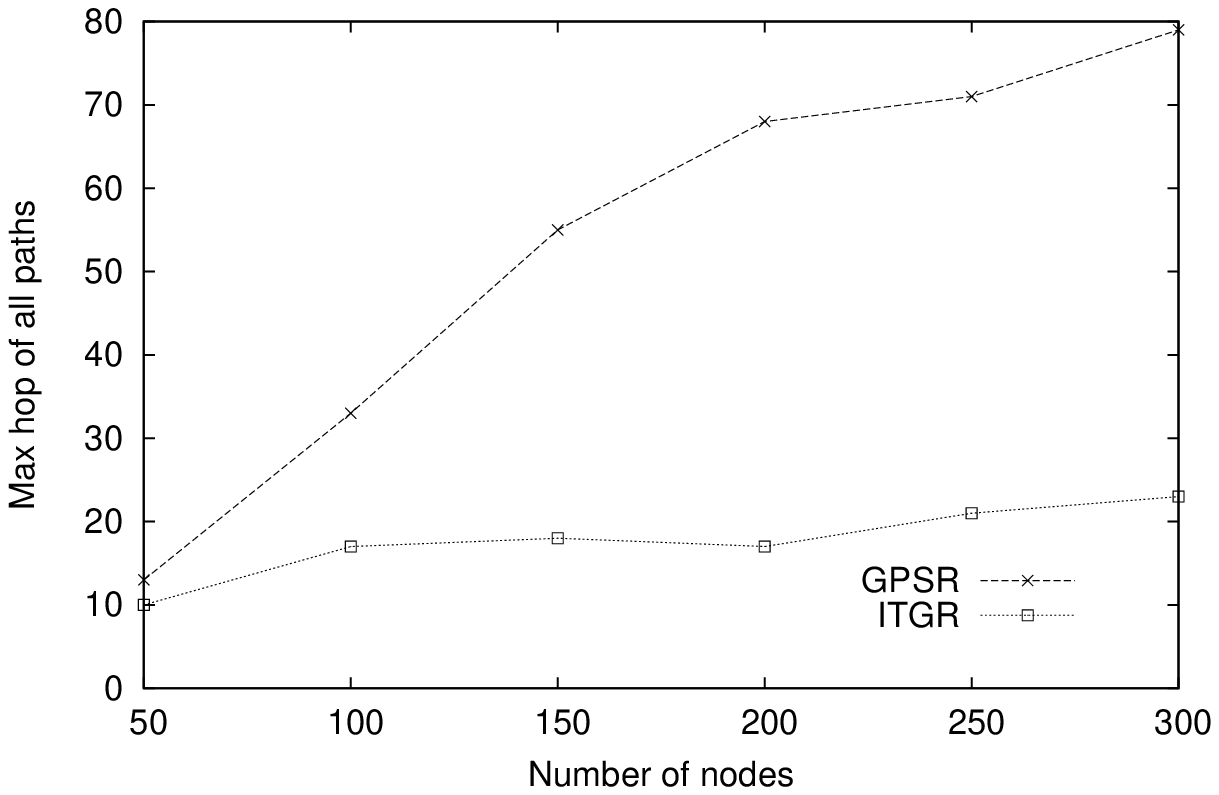}
\end{center}
\caption{The maximal number of hops of the paths}
\label{fig22}
\end{figure}

One benefit from ITGR is the reduction of the long detour path. To see the effect more clearly,
we are interested in observing the longest paths (measured either in length or in number of hops)
in ITGR and GPSR.
We compare the length of the longest paths generated
by ITGR and GPSR in Fig.~\ref{fig21}.
When there are 50 nodes in the networks, we do not see much difference. However,
when the number of nodes increases from 100 to 300, the length
of the longest path generated by GPSR also increases from 2 times to almost 5 times
the length of the longest path generated by ITGR.
In Fig.~\ref{fig22}, we compare the maximal number of hops of the paths in ITGR and GPSR.
We can see a similar pattern. When the number of nodes increases,
the difference between GPSR and ITGR becomes larger.
Hence ITGR can avoid most of the long detour paths resulted from GPSR.

ITGR is a hybrid protocol containing both proactive and reactive aspects. The proactive
operation is to save the $<$ $LocalMinimum$, $Landmark$ $>$ entries to
a local cache.
From our experiments, we find that the number of nodes that save the entries is not large,
relative to the number of all nodes in the networks.
Table~\ref{table3} shows that the number of nodes with cache entries
when the number of nodes in the networks changes
from 50 to 300. It  shows that the number of nodes with entries is about 10\%
of the total number of nodes in the networks.

\begin{table}[ht]
\caption{The average  number of nodes with cache entries and percentages over all nodes}
\centering          
\begin{tabular}{c c c c c c c}    
\hline\hline                        
network size & 50 & 100 & 150 & 200 & 250 & 300 \\ [0.5ex]  
\hline                      
number of nodes with entries &4&9&16&18&21&25 \\      
Percentage(\%)   &8.0&9.0&10.67&9.0&8.4&8.3\\  [1ex]        
\hline          
\end{tabular}
\label{table3}    
\end{table}

\begin{figure}[!htp]
\begin{center}
\includegraphics[width=9.0cm]{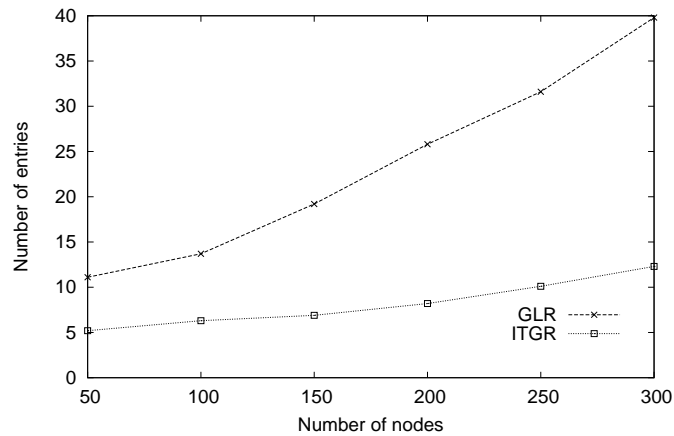}
\end{center}
\caption{The overheads of ITGR and GLR}
\label{overhead}
\end{figure}

Finally, we examine the control overhead of ITGR, by comparing it with GLR.
The control overhead is measured in term of the number of cache
entries saved in the nodes.
We calculate the number of cache entries stored at each node. The
overall overhead is the summation of these numbers.
Fig.~\ref{overhead}
shows
that the overhead of ITGR is much smaller than that of GLR.
When the number of nodes in the network increases from 50 to 300, the difference
in number of entries between the
two schemes becomes larger.
Because the entry of GLR is in the format of $<B_i, D_i>$,
GLR has to save an entry for
almost every destination node hidden behind a hole.
On the contrary, the entry of ITGR is in the form of $<P_i, B_i>$,
which can cover an area containing many destination nodes.

\begin{figure}[!hb]
\begin{center}
\includegraphics[width=9.0cm]{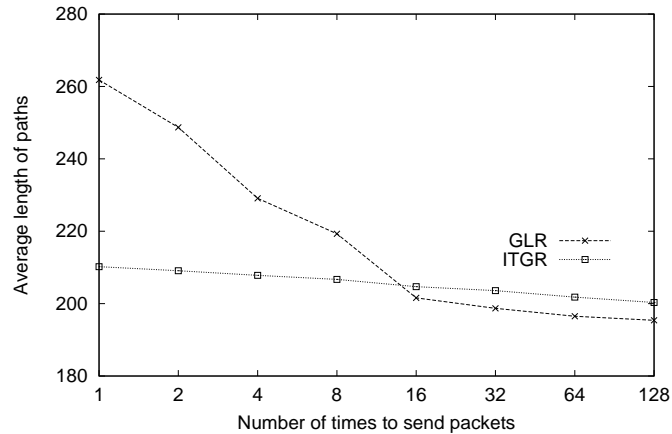}
\end{center}
\caption{The average length of paths of ITGR and GLR}
\label{exptestpath}
\end{figure}

To compare the performance of ITGR and GLR in terms of path length,
we randomly generate 100 networks with 150 nodes each.
In each network, 100 pairs
of source and destination nodes are randomly selected.
Since both schemes use previous experience to improve the performance of future transmissions,
sending to the same destination multiple times will get better results.
Therefore,
for each pair of nodes,
we present the results when
the source repeatedly sends a packet to the destination from once to 128 times.
The average length of paths generated by all the  100 pairs of nodes in all the 100 networks are reported in Fig.~\ref{exptestpath}.
The average length of paths
of GLR is a little shorter than that of ITGR only when the number of repeatedly sending times is larger
than 16, but not significantly.
When the number of the repeatedly sending times is less than 16, ITGR generates shorter paths than GLR.
This is because ITGR can improve the routing performance even if the source node has not sent a packet
to the same destination before.


\section{Conclusion}
\label{conclusion}

In this paper, we presented a new geographic routing approach called ITGR
in order to avoid the long detour path. It detects the destination areas
that might be shaded by the holes from previous routing experience.
Then it selects the landmarks as tentative targets to construct greedy sub-paths.
The approach can be used to avoid local minimum nodes.
We design the scheme in such a way that
a single detour path to a given destination
can be used to avoid the detour path to many destinations in the future.
We demonstrated a simple representation used for determining whether a node is in
the shaded area. We also developed a method to reduce the
overhead at nodes by combining multiple entries into one.
The simulations demonstrate that our approach can result in significant shorter
routing path and fewer hops than an existing geographic routing algorithm.


\bibliographystyle{elsarticle-num}
\bibliography{egbib}
~~~\\
~~~\\







\end{document}